\documentclass{llncs} 

\usepackage{latexsym}
\newcommand\eps\varepsilon

\title{RAM-Efficient External Memory Sorting\thanks{This paper will
    appear in the Proceedings of The 24th International Symposium on
    Algorithms and Computation, LNCS 8283, Springer, 2013.}}

\author{Lars Arge\inst{1}\thanks{Supported in part by the Danish National Research Foundation and the Danish National Advanced Technology Foundation.} and Mikkel Thorup\inst{2}\thanks{Supported in part by an Advanced Grant from the Danish Council for Independent Research under the Sapere Aude research career program.}}

\institute{MADALGO\thanks{Center for Massive Data Algorithmics---a
    center of the Danish National Research Foundation}, Aarhus
  University, Aarhus, Denmark \and 
  University of Copenhagen,\thanks{Part of this work was done while the author was at AT\&T Labs--Research.}
  Copenhagen, Denmark\vspace{-3ex}}

\begin{document}
\maketitle

\begin{abstract}
In recent years a large number of problems have been considered in
external memory models of computation, where the complexity measure is
the number of blocks of data that are moved between slow external
memory and fast internal memory (also called I/Os). In practice,
however, internal memory time often dominates the total running time
once I/O-efficiency has been obtained. In this paper we study
algorithms for fundamental problems that are simultaneously
I/O-efficient and internal memory efficient in the RAM model of
computation.\vspace{-3ex}


\end{abstract}

\section{Introduction} 

In the last two decades a large number of problems have been
considered in the external memory model of computation, where the
complexity measure is the number of blocks of elements that are moved
between external and internal memory. Such movements are also called
I/Os. The motivation behind the model is that random access to
external memory, such as disks, often is many orders of magnitude
slower than random access to internal memory; on the other hand, if
external memory is accessed sequentially in large enough blocks, then
the cost per element is small. In fact, disk systems are often
constructed such that the time spent on a block access is comparable
to the time needed to access each element in a block in internal
memory.

Although the goal of external memory algorithms is to minimize the
number of costly blocked accesses to external memory when processing
massive datasets, it is also clear from the above that if the internal
processing time per element in a block is large, then the practical
running time of an I/O-efficient algorithm is dominated by internal
processing time. Often I/O-efficient algorithms are in
fact not only efficient in terms of I/Os, but can also be shown to be
internal memory efficient in the comparison model. Still, in many
cases the practical running time of I/O-efficient algorithms is
dominated by the internal computation time. Thus both from a practical
and a theoretical point of view it is interesting to investigate how
internal-memory efficient algorithms can be obtained while
simultaneously ensuring that they are I/O-efficient. In this paper we
consider algorithms that are both I/O-efficient and efficient in the
RAM model in internal memory.

\paragraph{Previous results.}
We will be working in the standard external memory model of
computation, where $M$ is the number of elements that fit in main
memory and an I/O is the process of moving a block of $B$ consecutive
elements between external and internal
memory~\cite{aggarwal:input}. We assume that $N\geq 2M$, $M\geq 2B$
and $B\geq 2$. Computation can only be performed on elements in main
memory, and we will assume that each element consists of one word. We
will sometime assume the comparison model in internal memory, that is,
that the only computation we can do on elements are
comparisons. However, most of the time we will assume the RAM model in
internal memory. In particular, we will assume that we can use
elements for addressing, e.g. trivially implementing permuting in
linear time. Our algorithms will respect the standard so-called
\emph{indivisibility assumption}, which states that at any given time
during an algorithm the original $N$ input elements are stored
somewhere in external or internal memory. Our internal memory time
measure is simply the number of performed operations; note that this
includes the number of elements transferred between internal and
external memory.

Aggarwal and Vitter~\cite{aggarwal:input} described sorting algorithms
using $O(\frac{N}{B}\log_{M/B} \frac{N}{B})$ I/Os. One of these
algorithms, external merge-sort, is based on $\Theta(M/B)$-way
merging. First $O(N/M)$ sorted runs are formed by repeatedly sorting
$M$ elements in main memory, and then these runs are merged together
$\Theta(M/B)$ at a time to form longer runs. The process continues for
$O(\log_{M/B} \frac{N}{M})$ phases until one is left with one sorted
list. Since the initial run formation and each phase can be performed
in $O(N/B)$ I/Os, the algorithm uses $O(\frac{N}{B}\log_{M/B}
\frac{N}{B})$ I/Os. Another algorithm, external distribution-sort, is
based on $\Theta(\sqrt{M/B})$-way splitting. The $N$ input elements
are first split into $\Theta(\sqrt{M/B})$ sets of roughly equal size,
such that the elements in the first set are all smaller than the
elements in the second set, and so on. Each of the sets are then split
recursively. After $O(\log_{\sqrt{M/B}} \frac{N}{M})=O(\log_{M/B}
\frac{N}{M})$ split phases each set can be sorted in internal
memory. Although performing the split is somewhat complicated, each phase can still
be performed in $O(N/B)$ I/Os. Thus also this algorithm uses
$O(\frac{N}{B}\log_{M/B} \frac{N}{B})$ I/Os.

Aggarwal and Vitter~\cite{aggarwal:input} proved that external merge-
and distribution-sort are I/O-optimal when the comparison model is
used in internal memory, and in the following we will use
$sort_E(N)$ to denote the number of I/Os \emph{per block of
  elements} of these optimal algorithms, that is,
$sort_E(N)=O(\log_{M/B} \frac{N}{B})$ and external comparison model
sort takes $\Theta(\frac{N}{B} sort_E(N))$ I/Os. (As described below,
the I/O-efficient algorithms we design will move $O(N\cdot sort_E(N))$
elements between internal and external memory, so $O(sort_E(N))$ will
also be the per element internal memory cost of obtaining external
efficiency.) When no assumptions other than the indivisibility
assumption are made about internal memory computation (i.e. covering
our definition of the use of the RAM model in internal memory),
Aggarwal and Vitter~\cite{aggarwal:input} proved that permuting $N$
elements according to a given permutation requires
$\Omega(\min\{N,\frac{N}{B}sort_E(N)\})$ I/Os. Thus this is also a
lower bound for RAM model sorting. For all practical values of $N$,
$M$ and $B$ the bound is $\Omega(\frac{N}{B}sort_E(N))$.
Subsequently, a large number of I/O-efficient algorithms have
been developed.
Of particular relevance for this paper, several priority
queues have been developed where insert and deletemin operations can
be performed in $O(\frac{1}{B}sort_E(N))$
I/Os amortized~\cite{arge:buffer,fadel:external,kumar:improved}. The
structure by Arge~\cite{arge:buffer} is based on the so-called
buffer-tree technique, which uses $O(M/B)$-way splitting, whereas the
other structures also use $O(M/B)$-way merging.

In the RAM model the best known sorting
algorithm uses $O(N\log\log N)$ time~\cite{han:determistic}. Similar
to the I/O-case, we use $sort_I(N)=O(\log\log N)$ to denote the
\emph{per element} cost of the best known sorting algorithm. If
randomization is allowed then this can be improved to
$O(\sqrt{\log\log n})$ expected time~\cite{han:integer}. A priority
queue can also be implemented so that the cost per operation is
$O(sort_I(N))$~\cite{thorup:equivalance}.

\paragraph{Our results.}
In Section~\ref{sec:sorting} we first discuss how both external
merge-sort and external distribution-sort can be implemented to use
optimal $O(N\log N)$ time if the comparison model is used in internal
memory, by using an $O(N\log N)$ sorting algorithm and (in the
merge-sort case) an $O(\log N)$ priority queue. We also show how these
algorithms can relatively easily be modified to use $$O(N\cdot
(sort_I(N)+sort_I(M/B)\cdot sort_E(N))) \rm{~and}$$ $$O(N\cdot
(sort_I(N)+sort_I(M)\cdot sort_E(N)))$$ time, respectively, if the RAM
model is used in internal memory, by using an $O(N\cdot sort_I(N))$
sorting algorithm and an $O(sort_I(N))$
priority queue.

The question is of course if the above RAM model sorting algorithms
can be improved. In Section~\ref{sec:sorting} we discuss how it seems
hard to improve the running time of the merge-sort algorithm, since it
uses a priority queue in the merging step. By using a linear-time
internal-memory splitting algorithm, however, rather than an $O(N\cdot
sort_I(N))$ sorting algorithm, we manage to improve the running time
of external distribution-sort to $$O(N\cdot (sort_I(N)+sort_E(N))).$$ Our new
\emph{split-sort} algorithm still uses $O(\frac{N}{B}sort_E(N))$
I/Os. Note that for small values of $M/B$ the $N\cdot sort_E(N)$-term,
that is, the time spent on moving elements between internal and
external memory, dominates the internal time. Given the conventional wisdom that merging is
superior to splitting in external memory, it is also 
surprising that a distribution algorithm outperforms a
merging algorithm.

In Section~\ref{sec:p-queue} we develop an I/O-efficient RAM model
priority queue by modifying the buffer-tree based structure of
Arge~\cite{arge:buffer}. The main modification consists of
removing the need for sorting of $O(M)$ elements every time a
so-called buffer-emptying process is performed. The structure supports
insert and deletemin operations in $O(\frac{1}{B}sort_E(N))$ I/Os and
$O(sort_I(N)+sort_E(N))$ time. Thus it can be used to develop another
$O(\frac{N}{B}sort_E(N))$ I/O and $O(N\cdot (sort_I(N)+sort_E(N)))$
time sorting algorithm.

Finally, in Section~\ref{sec:lower} we show that when
$\frac{N}{B}sort_E(N)=o(N)$ (and our sorting algorithms are
I/O-optimal), any I/O-optimal sorting algorithm must
transfer a number of elements between internal and external memory
equal to $\Theta(B)$ times the number of I/Os it performs, that is, it
must transfer $\Omega(N\cdot sort_E(N))$ elements and thus also use
$\Omega(N\cdot sort_E(N))$ internal time. In fact, we show a
lower bound on the number of I/Os needed by an algorithm that
transfers $b\leq B$ elements on the average per I/O, significantly
extending the lower bound of Aggarwal and
Vitter~\cite{aggarwal:input}. The result implies that (in the
practically realistic case) when our split-sort and priority queue
sorting algorithms are I/O-optimal, they are in fact also CPU optimal
in the sense that their running time is the sum of an unavoidable term
and the time used by the best known RAM sorting algorithm. As mentioned above, the lower bound
also means that the time spent on moving elements between internal and
external memory resulting from the fact that we are considering
I/O-efficient algorithms can dominate the internal computation time,
that is, considering I/O-efficient algorithms implies that less
internal-memory efficient algorithms can be obtained than if not
considering I/O-efficiency. Furthermore, we show that when $B\leq M^{1-\eps}$ for some constant $\eps>0$ (the \emph{tall cache assumption}) the same $\Omega(N\cdot sort_E(N))$ number of transfers
are needed for any algorithm using less than $\eps N/4$ I/Os
(even if it is not I/O-optimal).

To summarize our contributions, we open up a new area of algorithms
that are both RAM-efficient and I/O-efficient. The area is interesting
from both a theoretical and practical point of view. We illustrate
that existing algorithms, in particular multiway merging based
algorithms, are not RAM-efficient, and develop a new sorting
algorithm that is both efficient in terms of I/O and RAM time, as
well as a priority queue that can be used in such an efficient
algorithm. We prove a lower bound that shows that our algorithms are
both I/O and internal-memory RAM model optimal. The lower bound
significantly extends the Aggarwal and Vitter lower
bound~\cite{aggarwal:input}, and shows that considering I/O-efficient
algorithms influences how efficient internal-memory algorithms can be
obtained.  
%


\section{Sorting}
\label{sec:sorting}


\paragraph{External merge-sort.}
In external merge-sort $\Theta(N/M)$ sorted runs are first formed by
repeatedly loading $M$ elements into main memory, sorting them, and
writing them back to external memory. In the first merge phase these
runs are merged together $\Theta(M/B)$ at a time to form longer
runs. The merging is continued for $O(\log_{M/B}
\frac{N}{M})=O(sort_E(N))$ merge phases until one is left with one
sorted run. It is easy to realize that $M/B$ runs can be merged
together in $O(N/B)$ I/Os: We simply load the first block of each of
the runs into main memory, find and output the $B$ smallest elements,
and continue this process while loading a new block from the relevant
run every time all elements in main memory from that particular run
have been output. Thus external merge-sort uses
$O(\frac{N}{B}\log_{M/B} \frac{N}{M})=O(\frac{N}{B}sort_E(N))$ I/Os.

In terms of internal computation time, the initial run formation can
trivially be performed in $O(N/M\cdot M\log M)=O(N\log M)$ time using
any $O(N\log N)$ internal sorting algorithm. Using an $O(\log (M/B))$
priority queue to hold the minimal element from each of the $M/B$ runs
during a merge, each of the $O(\log_{M/B} \frac{N}{M})$ merge phases
can be performed in $O(N\log \frac{M}{B})$ time. Thus external
merge-sort can be implemented to use $O(N\log M+\log_{M/B}
\frac{N}{M}\cdot N\log \frac{M}{B})=O(N\log M+N\log
\frac{N}{M})=O(N\log N)$ time, which is optimal in the comparison
model.

When the RAM model is used in internal memory, we can
improve the internal time by using a RAM-efficient
$O(M\cdot sort_I(M))$ algorithm in the run formation phase and by
replacing the $O(\log (M/B))$ priority queue with an $O(sort_I(M/B))$ time
priority queue~\cite{thorup:equivalance}. This leads to an
$O(N\cdot(sort_I(M)+sort_I(M/B)\cdot sort_E(N))$ algorithm. There seems
no way of avoiding the extra $sort_I(M/B)$-term, since that
would require an $O(1)$ priority queue.

\paragraph{External distribution-sort.}
In external distribution-sort the input set of $N$ elements is first
split into $\sqrt{M/B}$ sets $X_0, X_1,\dots, X_{\sqrt{M/B}-1}$
defined by $s=\sqrt{M/B}-1$ split elements $x_1<x_2<\dots<x_s$,
such that all elements in $X_0$ are smaller than $x_1$, all elements
in $X_{\sqrt{M/B}-1}$ are larger than or equal to $x_s$, and such that
for $1\leq i\leq \sqrt{M/B}-2$ all elements in $X_i$ are larger than
or equal to $x_i$ and smaller than $x_{i+1}$. Each of these sets is
recursively split until each set is smaller than $M$ (and larger than
$M/(M/B)=B$) and can be sorted in internal memory. If the $s$
split elements are chosen such that $|X_i|=O(N/s)$ then there are
$O(\log_s \frac{N}{B})=O(\log_{M/B} \frac{N}{B})=O(sort_E(N))$ split
phases. Aggarwal and Vitter~\cite{aggarwal:input} showed how to
compute a set of $s$ split elements with this property in $O(N/B)$
I/Os. Since the actual split of the elements according to the
split elements can also be performed in $O(N/B)$ I/Os (just like merging of
$M/B$ sorted runs), the total number of I/Os needed by
distribution-sort is $O(\frac{N}{B}sort_E(N))$.

Ignoring the split element computation it is easy to implement
external distribution-sort to use $O(N\log N)$ internal 
time in the comparison model: During a split we simply hold the split
elements in main memory and perform a binary search among them with
each input element to determine to which set $X_i$ the element should
go. Thus each of the $O(\log_{M/B} \frac{N}{B})$ split phases uses
$O(N\log \sqrt{M/B})$ time. Similarly, at the end of the recursion we
sort $O(N/M)$ memory loads using $O(N\log M)$ time in total. The split
element computation algorithm of Aggarwal and
Vitter~\cite{aggarwal:input}, or rather its analysis, is somewhat
complicated. Still it is easy to realize that it also works in
$O(N\log M)$ time as required to obtain an $O(N\log N)$ time algorithm
in total. The algorithm works by loading the $N$ elements a memory
load at a time, sorting them and picking every $\sqrt{M/B}/4$'th
element in the sorted order. This obviously requires $O(N/M\cdot
M\log M)=O(N\log M)$ time and results in a set of $4N/\sqrt{M/B}$
elements. Finally, a linear I/O and time algorithm is used
$\sqrt{M/B}$ times on this set of elements to obtain the split
elements, thus using $O(N)$ additional time.

If we use a RAM sorting algorithm to sort the memory loads at the end
of the split recursion, the running time of this part of the algorithm
is reduced to $O(N\cdot sort_I(M))$. Similarly, we can use the RAM
sorting algorithm in the split element computation algorithm,
resulting in an $O(N\cdot sort_I(M))$ algorithm and consequently a
$sort_I(M)$-term in the total running time. Finally, in order to
avoid the binary search over $\sqrt{M/B}$ split elements in the
actual split algorithm, we can modify it to use sorting instead: To
split $N$ elements among $s$ splitting elements stored in $s/B$ blocks
in main memory, we allocate a buffer of one block in main memory for
each of the $s+1$ output sets. Thus in total we require
$s/B+(s+1)B<M/2$ of the main memory for split elements and
buffers. Next we repeatedly bring $M/2$ elements onto main memory,
sort them, and distribute them to the $s+1$ buffers, while outputting
the $B$ elements in a buffer when it runs full. Thus this process
requires $O(N\cdot sort_I(M))$ time and $O(N/B)$ I/Os like the split
element finding algorithm. Overall this leads to an $O(N\cdot
(sort_I(M)+sort_I(M)\cdot sort_E(N)))$ time algorithm.

\paragraph{Split-sort.}
While it seems hard to improve the RAM running time of
the external merge-sort algorithm, we can actually modify the external
distribution-sort algorithm further and obtain an algorithm that in
most cases is optimal both in terms of I/O and time. This
\emph{split-sort} algorithm basically works like the distribution-sort
algorithm with the split algorithm modification described
above. However, we need to modify the algorithm further in order to
avoid the $sort_I(M)$-term in the time bound that appears due to the
repeated sorting of $O(M)$ elements in the split element finding
algorithm, as well as in the actual split algorithm.

First of all, instead of sorting each batch of $M/2$ elements in the
split algorithm to split them over $s=\sqrt{M/B}-1<\sqrt{M/2}$ split
elements, we use a previous result that shows that we can actually
perform the split in linear time.

\begin{lemma}[Han and Thorup \cite{han:integer}]
\label{lem:split}
  In the RAM model $N$ elements can be split over $N^{1-\eps}$ split
  elements in linear time and space for any constant $\eps>0$.
\end{lemma}

Secondly, in order to avoid the sorting in the split element finding
algorithm of Aggarwal and Vitter~\cite{aggarwal:input}, we design a
new algorithm that finds the split elements on-line as part of the actual
split algorithm, that is, we start the splitting with no split
elements at all and gradually add at most $s=\sqrt{M/B}-1$ split
elements one at a time. An online split strategy was previously used by Frigo et al~\cite{frigo:oblivious} in a cache-oblivious algorithm setting. More precisely, our algorithm works as
follows. To split $N$ input elements we, as previously, repeatedly
bring $M/2$ elements onto main memory, distribute them to buffers
using the current split elements and Lemma~\ref{lem:split}, while
outputting the $B$ elements in a buffer when it runs full. However,
during the process we keep track of how many elements are output to
each subset. If the number of elements in a subset $X_i$ becomes
$2N/s$ we pause the split algorithm, compute the median of $X_i$ and
add it to the set of splitters, and split $X_i$ at the median element
into two sets of size $N/s$. Then we continue the splitting algorithm.

It is easy to see that the above splitting process results in at most
$s+1$ subsets containing between $N/s$ and $2N/s-1$ elements each,
since a set is split when it has $2N/s$ elements and each new set
(defined by a new split element) contains at least $N/s$ elements. The
actual median computation and the split of $X_i$ can be performed in
$O(|X_i|)=O(N/s)$ time and $O(|X_i|/B)=O(N/sB)$
I/Os~\cite{aggarwal:input}. Thus if we charge this cost to the at
least $N/s$ elements that were inserted in $X_i$ since it was created, each
element is charged $O(1)$ time and $O(1/B)$ I/Os. Thus each
distribution phase is performed in linear time and $O(N/B)$ I/Os,
leading to an $O(N\cdot (sort_I(M)+sort_E(N)))$ time algorithm.

\begin{theorem}
  The split-sort algorithm can be used to sort $N$ elements in
  $O(N\cdot (sort_I(M)+sort_E(N)))$ time and $O(\frac{N}{B}sort_E(N))$
  I/Os.
\end{theorem}

\noindent
\emph{Remarks.} Since $sort_I(M)+sort_E(N)\geq sort_I(N)$ our
split-sort algorithm uses $\Omega(N\cdot sort_I(N))$ time. In
Section~\ref{sec:lower} we prove that the algorithm in some sense is
optimal both in terms of I/O and time. Furthermore, we believe that
the algorithm is simple enough to be of practical interest.


\section{Priority queue}
\label{sec:p-queue}

In this section we discuss how to implement an I/O- and RAM-efficient
priority queue by modifying the I/O-efficient buffer tree priority queue~\cite{arge:buffer}.


\paragraph{Structure.}
Our external priority queues consists of a fanout $\sqrt{M/B}$
B-tree~\cite{cormer:ubiquitous} $T$ over $O(N/M)$ leaves containing
between $M/2$ and $M$ elements each. In such a tree, all leaves are on
the same level and each node (except the root) has fan-out between
$\frac{1}{2}\sqrt{M/B}$ and $\sqrt{M/B}$ and contains at most $\sqrt{M/B}$
splitting elements defining the element ranges of its children. Thus
$T$ has height $O(\log_{\sqrt{M/B}} \frac{N}{M})=O(sort_E(N))$. To
support insertions efficiently in a ``lazy'' manner, each internal
node is augmented with a \emph{buffer} of size $M$ and an
\emph{insertion buffer} of size at most $B$ is maintained in internal
memory. To support deletemin operations efficiently, a RAM-efficient
priority queue~\cite{thorup:equivalance} supporting both deletemin and
deletemax,\footnote{A priority queue supporting both deletemin and
  deletemax can easily be obtained using two priority queues
  supporting deletemin and delete as the one by
  Thorup~\cite{thorup:equivalance}.} called the \emph{mini-queue}, is
maintained in main memory containing the up to $M/2$ smallest elements
in the priority queue.

\paragraph{Insertion.}
To perform an insertion we first check if the element to be inserted
is smaller than the maximal element in the mini-queue, in which case
we insert the new element in the mini-queue and continue the insertion
process with the currently maximal element in the mini-queue. Next we
insert the element to be inserted in the insertion buffer. When we
have collected $B$ elements in the insertion buffer we insert them in the
buffer of the root. If this buffer now contains more than $M/2$
elements we perform a \emph{buffer-emptying process} on it,
``pushing'' elements in the buffer one level down to buffers on the
next level of $T$: We load the $M/2$ oldest elements into main memory
along with the less than $\sqrt{M/B}$ splitting elements, distribute
the elements among the splitting elements, and finally output them
to the buffers of the relevant children. Since the splitting and
buffer elements fit in memory and the buffer elements are distributed
to $\sqrt{M/B}$ buffers one level down, the buffer-emptying process is
performed in $O(M/B)$ I/Os. Since we distribute $M/2$ elements using
$\sqrt{M/B}$ splitters the process can be performed in $O(M)$ time
(Lemma~\ref{lem:split}).
After emptying the buffer of the root some of the nodes on the next
level may contain more than $M/2$ elements. If they do we perform
recursive buffer-emptying processes on these nodes. Note that this way
buffers will never contain more than $M$ elements.
When (between $1$ and $M/2$) elements are pushed down to a leaf (when
performing a buffer-emptying process on its parent) resulting in the
leaf containing more than $M$ (and less than $3M/2$) elements we split
it into two leaves containing between $M/2$ and $3M/4$ elements
each. We can easily do so in $O(M/B)$ I/Os and $O(M)$
time~\cite{aggarwal:input}. As a result of the split the parent node
$v$ gains a child, that is, a new leaf is inserted. If needed, $T$ is
then balanced using node splits as a normal B-tree, that is, if the
parent node now has $\sqrt{M/B}$ children it is split into two nodes
with $1/2\sqrt{M/B}$ children each, while also distributing the
elements in $v$'s buffer among the two new nodes. This can easily be
accomplished in $O(M/B)$ I/Os and $M$ time. The rebalancing may
propagate up along the path to the root (when the root splits a new
root with two children is constructed).

During buffer-emptying processes we push $\Theta(M)$ elements one
level down the tree using $O(M/B)$ I/Os and $O(M)$ time. Thus each
element inserted in the root buffer pays $O(1/B)$ I/Os and $O(1)$ time
amortized, or $O(\frac{1}{B}\log_{M/B}
\frac{N}{B})=O(\frac{1}{B}sort_E(N))$ I/Os and $O(\log_{M/B}
\frac{N}{B})=O(sort_E(N))$ time amortized on buffer-emptying processes
on a root-leaf path. When a leaf splits we may use $O(M/B)$ I/Os and
$O(M)$ time in each node of a leaf-root path of length
$O(sort_E(N))$. Amortizing among the at least $M/4$ elements that were
inserted in the leaf since it was created, each element is charged and
additional $O(\frac{1}{B}sort_E(N))$ I/Os and $O(sort_E(N))$ time on
insertion in the root buffer. Since insertion of an element in the
root buffer is always triggered by an insertion operation, we can
charge the $O(\frac{1}{B}sort_E(N))$ I/Os and $O(sort_E(N))$ time cost
to the insertion operation.

\paragraph{Deletemin.}
To perform a deletemin operation we first check if the mini-queue
contains any elements. If it does we simply perform a deletemin
operation on it and return the retrieved element using $O(sort_I(M))$
time and no I/Os. Otherwise we perform buffer-emptying processes on
all nodes on the leftmost path in $T$ starting at the root and moving
towards the leftmost leaf. After this the buffers on the leftmost path
are all empty and the smallest elements in the structure are stored in
the leftmost leaf. We load the between $M/2$ and $M$ elements in the
leaf into main memory, sort them and remove the smallest $M/2$
elements and insert them in the mini-queue in internal memory. If this
results in the leaf having less than $M/2$ elements we insert the
elements in a sibling and delete the leaf. If the sibling now has
more than $M$ elements we split it. As a result of this the parent
node $v$ may lose a child. If needed $T$ is then rebalanced using
node fusions as a normal B-tree, that is, if $v$ now has
$1/2\sqrt{M/B}$ children it is fused with its sibling (possibly
followed by a split). As with splits after insertion of a new leaf,
the rebalancing may propagate up along the path to the root (when the
root only has one leaf left it is removed). Note that no buffer
merging is needed since the buffers on the leftmost path are all
empty.

If buffer-emptying processes are needed during a deletemin operation
we spend $O(\frac{M}{B}\log_{M/B}
\frac{N}{B})=O(\frac{M}{B}sort_E(N))$ I/Os and $O(M\log_{M/B}
\frac{N}{B})=O(M\cdot sort_E(N))$ time on such processes that are not
paid by buffers running full (containing more than $M/2$
elements). We also use $O(M/B)$ I/Os and $O(M\cdot sort_I(M))$ time to
load and sort the leftmost leaf, and another $O(M\cdot sort_I(M))$
time is used to insert the $M/2$ smallest elements in the
mini-queue. Then we may spend $(M/B)$ I/Os and $O(M)$ time on each of
at most $O(\log_{M/B} \frac{N}{B})$ nodes on the leftmost path that
need to be fused or split. Altogether the filling up of the mini-queue
requires $O(\frac{M}{B}sort_E(N))$ I/Os and $O(M\cdot
(sort_I(M)+sort_E(N)))$ time. Since we only fill up the mini-queue
when $M/2$ deletemin operations have been performed since the last
fill up, we can amortize this cost over these $M/2$ deletemin
operations such that each deletemin is charged
$O(\frac{1}{B}sort_E(N))$ I/Os and $O(sort_E(N)+sort_I(M))$ time.

\begin{theorem} 
  There exists a priority queue supporting an insert operation in
  $O(\frac{1}{B}sort_E(N))$ I/Os and $O(sort_E(N))$ time
  amortized and a deletemin operation in $O(\frac{1}{B}sort_E(N))$
  I/Os and $O(sort_I(M)+sort_E(N))$ time amortized.
\end{theorem}

\noindent
\emph{Remarks.} Our priority queue obviously can be used in a simple
$O(\frac{N}{B}sort_E(N))$ I/O and $O(N\cdot (sort_I(M)+sort_E(N)))$
time sorting algorithm. Note that it is essential that a
buffer-emptying process does not require sorting of the elements in
the buffer. In normal buffer-trees~\cite{arge:buffer} such a sorting
is indeed performed, mainly to be able to support deletions and
(batched) rangesearch operations efficiently. Using a more elaborate
buffer-emptying process we can also support deletions without the need
for sorting of buffer elements.

\section{Lower bound}
\label{sec:lower}

%

Assume that $\frac{N}{B}sort_E(N)=o(N)$ and for simplicity also that
$B$ divides $N$. Recall that under the indivisibility assumption we
assume the RAM model in internal memory but require that at any time
during an algorithm the original $N$ elements are stored somewhere in memory; we allow copying of the original
elements. The internal memory contains at most $M$ elements and the
external memory is divided into $N$ \emph{blocks} of $B$ elements
each; we only need to consider $N$ blocks, since we are considering
algorithms doing less than $N$ I/Os. During an algorithm, we let $X$
denote the set of original elements (including copies) in internal
memory and $Y_i$ the set of original elements (including copies) in
the $i$'th block; an I/O transfers \emph{up to} $B$ elements between
an $Y_i$ and $X$. Note that in terms of CPU time, an I/O can cost
anywhere between $1$ and $B$ (transfers).

In the external memory permuting problem, we are given $N$ elements in
the first $N/B$ blocks and want to rearrange them according to a given
permutation; since we can always rearrange the elements within the
$N/B$ blocks in $O(N/B)$ I/Os, a permutation is simply given as an
assignment of elements to blocks (i.e. we ignore the order of the
elements within a block). In other words, we start with a distribution
of $N$ elements in $X, Y_1, Y_2,\dots Y_N$ such that $|Y_1|=|Y_2|=\dots
=|Y_{N/B}|=B$ and $X=Y_{(N/B)+1}=Y_{(N/B)+2}=\dots =Y_N=\emptyset$,
and should produce another given distribution of the same elements
such that $|Y_1|=|Y_2|=\dots =|Y_{N/B}|=B$ and
$X=Y_{(N/B)+1}=Y_{(N/B)+2}=\dots =Y_N=\emptyset$.

To show that any permutation algorithm that performs
$O(\frac{N}{B}sort_E(N))$ I/Os has to transfer $\Omega(N\cdot
sort_E(N))$ elements between internal and external memory, we first
note that at any given time during a permutation algorithm we can
identify a distribution (or more) of the original $N$ elements (or
copies of them) in $X, Y_1, Y_2,\dots Y_N$. We
then first want to bound the number of distributions that can be
created using $T$ I/Os, given that $b_i$, $1\leq i\leq T$, is the
number of elements transferred in the $i$'th I/O; any correct
permutation algorithm needs to be able to create at least
$\frac{N!}{B!^{N/B}}=\Omega((N/B)^N)$ distributions.

Consider the $i$'th I/O. There are at most $N$ possible choices for
the block $Y_j$ involved in the I/O; the I/O either transfers $b_i\leq
B$ elements from $X$ to $Y_j$ or from $Y_j$ to $X$. In the first case
there are at most ${M \choose b_i}$ ways of choosing the $b_i$
elements, and each element is either moved or copied. In the second
case there are at most most ${B \choose b_i}$ ways of choosing the
elements to move or copy. Thus the I/O can at most increase the number
of distributions that can be created by a factor of \[N\cdot \left({M
  \choose b_i}+{B \choose b_i}\right)\cdot
2^{b_i}<N(2eM/b_i)^{2b_i}.\]
\noindent
Now the $T$ I/Os can thus at most create $\prod_{i=1}^T
N(2eM/b_i)^{2b_i}$ distributions. That this number is bounded by
$\left(N(2eM/b)^{2b}\right)^T$, where $b$ is the average of the
$b_i$'s, can be seen by just considering two values $b_1$ and $b_2$
with average $b$. In this case we have
\[N(2eM/b_1)^{2b_1}\cdot N(2eM/b_2)^{2b_2}\leq
\frac{N^2(2eM)^{2(b_1+b_2)}}{b^{2(b_1+b_2)}}\leq
\left(N(2eM/b)^{2b}\right)^2.\]

Next we consider the number of distributions that can be created using
$T$ I/Os for all possible values of $b_i$, $1\leq i\leq T$, with a
given average $b$. This can trivially be bounded by multiplying the
above bound by $B^T$ (since this is a bound on the total number of
possible sequences $b_1, b_2,\dots ,b_T$). Thus the number of
distributions is bounded by $B^T\left(N(2eM/b)^{2b}\right)^T=
((BN)(2eM/b)^{2b})^T.$ Since any permutation algorithm needs to be
able to create $\Omega((N/B)^N)$ distributions, we get the following
lower bound on the number of I/Os $T(b)$ needed by an algorithm that
transfers $b\leq B$ elements on the average per
I/O: \[T(b)=\Omega\left(\frac{N\log (N/B)}{\log N+b\log
  (M/b)}\right).\]

\noindent
Now $T(B)=\Omega(\min\{N,\frac{N}{B}sort_E(N)\})$ corresponds to the
lower bound proved by Aggarwal and Vitter~\cite{aggarwal:input}. Thus
when $\frac{N}{B}sort_E(N)=o(N)$ we get
$T(B)=\Omega(\frac{N}{B}sort_E(N))=\Omega\left(\frac{N\log
  (N/B)}{B\log (M/B)}\right).$ Since $1\leq b\leq B\leq M/2$, we have
$T(b)=\omega(T(B))$ for $b=o(B)$. Thus any algorithm performing optimal
$O(\frac{N}{B}sort_E(N))$ I/Os must transfer $\Omega(N\cdot
sort_E(N))$ elements between internal and external memory.

Reconsider the above analysis under the tall
cache assumption $B\leq M^{1-\eps}$ for some constant $\eps >0$. In
this case, we have that the number of distributions any permutation
algorithm needs to be able to create is
$\Omega((N/B)^N)=\Omega(N^{\eps N})$. Above we
proved that with $T$ I/Os transferring an average number of $b$ keys
an algorithm can create at most $(BN(2eM/b)^{2b})^T< N^{2T}M^{2bT}$
distributions. Thus we have $M^{2bT}\geq N^{\eps N-2T}$. For $T<\eps
N/4$, we get $M^{2bT}\geq N^{\eps N/2}$ and thus that the number of
transferred elements $bT$ is $\Omega(N \log_M N)$. Since the tall
cache assumption implies that $\log (N/B)=\Theta(\log N)$ and $\log
(M/B)=\Theta(\log M)$ we have that $N\log_M N=\Theta(N\log_{M/B}
(N/B))=\Theta(N\cdot sort_E(N))$. Thus any algorithm using less than
$\eps N/4$ I/Os must transfer $\Omega(N\cdot sort_E(N))$ elements
between internal and external memory.

\begin{theorem}
  When $B\leq \frac{1}{2}M$ and $\frac{N}{B}sort_E(N)=o(N)$, any
  I/O-optimal permuting algorithm must transfer $\Omega(N\cdot
  sort_E(N))$ elements between internal and external memory under the
 indivisibility assumption.

  When $B\leq M^{1-\eps}$ for some constant $\eps>0$ any, permuting
  algorithm using less than $\eps N/4$ I/Os must transfer
  $\Omega(N\cdot sort_E(N))$ elements between internal and external
  memory under the indivisibility assumption.
\end{theorem}

\noindent
\emph{Remark.} The above means that in practice where
$\frac{N}{B}sort_E(N)=o(N)$ our $O(\frac{N}{B}sort_E(N))$ I/O and
$O(N\cdot (sort_I(N)+sort_E(N))$ time split-sort and priority queue
sort algorithms are not only I/O-optimal but also CPU optimal in the
sense that their running time is the sum of an unavoidable term and
the time used by the best known RAM sorting algorithm.


\begin{thebibliography}{1}

\bibitem{aggarwal:input}
A.~Aggarwal and J.~S. Vitter.
\newblock The {I}nput/{O}utput complexity of sorting and related problems.
\newblock {\em Communications of the ACM}, 31(9):1116--1127, 1988.

\bibitem{arge:buffer}
L.~Arge.
\newblock The buffer tree: A technique for designing batched external data
  structures.
\newblock {\em Algorithmica}, 37(1):1--24, 2003.

\bibitem{cormer:ubiquitous}
D.~Comer.
\newblock The ubiquitous {B}-tree.
\newblock {\em ACM Computing Surveys}, 11(2):121--137, 1979.

\bibitem{fadel:external}
R.~Fadel, K.~V. Jakobsen, J.~Katajainen, and J.~Teuhola.
\newblock Heaps and heapsort on secondary storage.
\newblock {\em Theoretical Computer Science}, 220(2):345--362, 1999.

\bibitem{frigo:oblivious}
M.~Frigo, C.~E. Leiserson, H.~Prokop, and S.~Ramachandran.
\newblock Cache-oblivious algorithms.
\newblock In {\em Proc. IEEE Symposium on Foundations of Computer Science},
  pages 285--298, 1999.

\bibitem{han:determistic}
Y.~Han.
\newblock Deterministic sorting in {$O(n \log\log n)$} time and linear space.
\newblock In {\em Proc. ACM Symposium on Theory of Computation}, pages
  602--608, 2002.

\bibitem{han:integer}
Y.~Han and M.~Thorup.
\newblock Integer sorting in {$O(n\sqrt{\log\log n})$} expected time and linear
  space.
\newblock In {\em Proc. IEEE Symposium on Foundations of Computer Science},
  pages 135--144, 2002.

\bibitem{kumar:improved}
V.~Kumar and E.~Schwabe.
\newblock Improved algorithms and data structures for solving graph problems in
  external memory.
\newblock In {\em Proc. IEEE Symp. on Parallel and Distributed Processing},
  pages 169--177, 1996.

\bibitem{thorup:equivalance}
M.~Thorup.
\newblock Equivalence between priority queues and sorting.
\newblock {\em Journal of the ACM}, 54(6):Article 28, 2007.

\end{thebibliography}
\end{document}